\documentclass[sigconf, 9pt, anonymous=false]{acmart}

\let\orgautoref\autoref
\renewcommand{\autoref}
{\def\sectionautorefname{Section}%
\def\subsectionautorefname{Section}%
\def\subsubsectionautorefname{Section}%
\orgautoref}

\usepackage{pifont}

\usepackage{xspace}
\newcommand{\etal}{\textit{et al.}}
\newcommand{\eg}{\textit{e.g.,}~}
\newcommand{\ie}{\textit{i.e.,}~}

\newcommand{\one}{({\em i})\xspace}
\newcommand{\two}{({\em ii})\xspace}
\newcommand{\three}{({\em iii})\xspace}
\newcommand{\four}{({\em iv})\xspace}
\newcommand{\five}{({\em v})\xspace}
\newcommand{\six}{({\em vi})\xspace}

\newcommand{\client}{LwM2M client}
\newcommand{\server}{LwM2M server}

\newcommand{\bserver}{LwM2M bootstrap-server}
\newcommand{\hclient}{hosting client}
\newcommand{\rclient}{requesting client}

\makeatletter
\renewcommand{\paragraph}[1]{\vspace*{0.03in}\noindent{\bf #1.}\hspace{0.25ex \@plus1ex \@minus.2ex}}

\newcommand{\paragraphS}[1]{\vspace*{0.03in}\noindent{\bf #1}\hspace{0.25ex \@plus1ex \@minus.2ex}}

\usepackage{xcolor}
\definecolor{colorOrange}{HTML}{fdb757}
\definecolor{colorDarkGray}{HTML}{333333}

\newcommand*\circledOrange[1]{\tikz[baseline=(char.base)]{
        \node[shape=circle,draw,inner sep=0.5pt,fill=colorOrange,text=colorDarkGray,font=\small] (char) {\bfseries{\textsf{#1}}};}}

\makeatother

\acmYear{2022}\copyrightyear{2022}
\setcopyright{rightsretained}
\acmConference[IPSN '22]{International Conference on Information Processing in Sensor Networks}{May 4--6, 2022}{Milan, Italy}
\acmBooktitle{International Conference on Information Processing in Sensor Networks (IPSN '22), May 4--6, 2022, Milan, Italy}
\acmPrice{15.00}

\usepackage{algorithmic}
\usepackage{graphicx}
\usepackage{textcomp}
\usepackage{xcolor}
\usepackage{subcaption}
\usepackage{hyperref}
\usepackage{booktabs}
\usepackage{makecell}
\usepackage{arydshln}
\newcommand*\dhline{\hdashline[.3pt/2pt]}

\usepackage{tikz}
\usepackage{tabularx}
\usepackage{array}
\usepackage{fontawesome}
\usepackage{rotating}
\usepackage[export]{adjustbox}
\usepackage{dblfloatfix}

\usepackage[absolute,showboxes]{textpos}

\newcolumntype{b}{X}
\newcolumntype{m}{>{\hsize=.5\hsize}X}
\newcolumntype{s}{>{\hsize=.25\hsize}X}
\newcolumntype{t}{>{\hsize=.125\hsize}X}
\newcolumntype{n}{>{\hsize=.05\hsize}X}

\newenvironment{Turn}{\begin{turn}{45}\begin{minipage}{1.5cm}\raggedright}
  {\end{minipage}\end{turn}}

\begin{document}

\renewcommand\footnotetextcopyrightpermission[1]{}

\settopmatter{printacmref=false,printccs=false,printfolios=false}

\title[Secure and Authorized Client-to-Client Communication for LwM2M]{Secure and Authorized Client-to-Client\\Communication for LwM2M}

\author{Leandro Lanzieri}
\affiliation{%
  \institution{HAW Hamburg}
}
\email{leandro.lanzieri@haw-hamburg.de}

\author{Peter Kietzmann}
\affiliation{%
  \institution{HAW Hamburg}
}
\email{peter.kietzmann@haw-hamburg.de}

\author{Thomas C. Schmidt}
\affiliation{%
  \institution{HAW Hamburg}
}
\email{t.schmidt@haw-hamburg.de}

\author{Matthias W{\"a}hlisch}
\affiliation{%
  \institution{Freie Universit{\"a}t Berlin}
}
\email{m.waehlisch@fu-berlin.de}

\renewcommand{\shortauthors}{Leandro Lanzieri, et al.}

\begin{abstract}
Constrained devices on the Internet of Things (IoT) continuously produce and consume data.
 LwM2M manages millions of these devices in a server-centric architecture,
 which challenges edge networks with expensive uplinks and time-sensitive use cases.
In this paper, we contribute two LwM2M extensions to enable client-to-client~(C2C) communication:
\one an authorization mechanism for clients,
and \two an extended management interface to allow secure C2C access to resources.
We analyse the security properties of the proposed extensions and show that they are compliant with LwM2M security requirements.
Our performance evaluation on off-the-shelf IoT hardware reveals that C2C communication outperforms server-centric deployments.
First, LwM2M deployments with edge C2C communication yield a $\approx$ 90\% faster notification delivery and $\approx 8 $ times higher throughput compared to common server-centric scenarios, while keeping a small memory overhead of $\approx$ 8\%.
Second, in server-centric communication, the delivery rate degrades when resource update intervals drop below 100 ms.
\end{abstract}



\maketitle

\setlength{\TPHorizModule}{\paperwidth}
\setlength{\TPVertModule}{\paperheight}
\TPMargin{5pt}
\begin{textblock}{0.8}(0.1,0.02)
     \noindent
     \footnotesize
     If you cite this paper, please use the IPSN reference:
     L. Lanzieri, P. Kietzmann, T. C. Schmidt, M. W\"ahlisch. Secure and Authorized Client-to-Client Communication for LwM2M. In \emph{Proc. of IPSN}, IEEE, 2022.
\end{textblock}

\section{Introduction}

%

\begin{figure}[t]
    \includegraphics[width=0.5\textwidth]{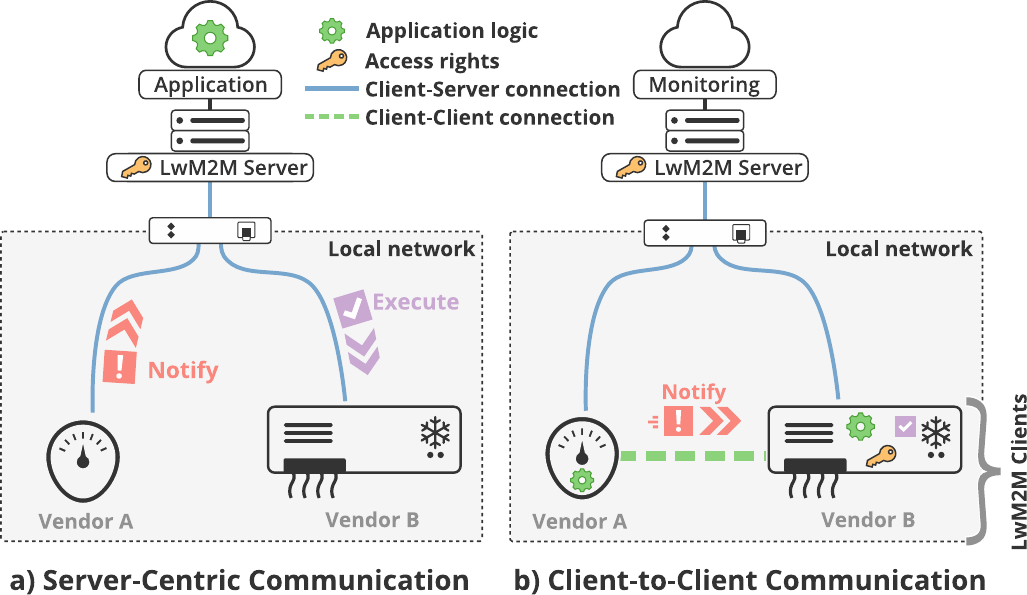}
    \caption{Different LwM2M deployment models. This paper introduces b) client-to-client communication.}
    \label{fig:deployment}
\end{figure}

The constant expansion of the Internet of Things (IoT) led to an increased deployment of proprietary ecosystems to interconnect resource constrained ``things'' via the global Internet.
Interoperability between an ever-increasing number of devices and vendors becomes paramount to avoid incompatibility silos.
Lightweight Machine to Machine (LwM2M) \cite{oma-lwm2m-core-12} is a widely deployed protocol that provides device management features, service enablement, and interoperability across vendors,
by defining an interaction model between LwM2M servers and clients, which operate on a uniform resource model.

In addition, the need for edge computing in IoT deployments~\cite{draft-irtf-t2trg-iot-edge} is rising,
driven by high data volume and constraints such as intermittent uplink connection to the server and low latency requirements.
In these scenarios, autonomous devices, executing distributed application logic are preferred.
LwM2M client-to-client (C2C) communication enables this, by allowing edge devices to perform operations directly, while dynamic resource discovery facilitates application logic specification on runtime without human intervention.
This requires a mechanism to dynamically distribute credentials and access rights to use resources of other clients securely.
LwM2M lacks such a direct client communication because it only allows servers to initiate transactions. Information always flows through servers, see \autoref{fig:deployment}a).

This work contributes to the research agenda of ``building an open, scalable, and secure Internet of Things'' that is accessible to all parties via standards. In particular, 
we fill a design gap by introducing two extensions to the LwM2M core specification and providing an open source implementation on RIOT~\cite{bghkl-rosos-18}.
Our proposal enables a secure and authorized communication regime between clients, see \autoref{fig:deployment}b).
In detail, we make the following contributions:
\begin{enumerate}
\item a third party authorization mechanism
\begin{anonsuppress}
\cite{lksw-tpalc-21}
\end{anonsuppress}
that allows clients to dynamically request servers to gain credentials and access rights to resources hosted by other clients.

\item new LwM2M objects and the extension of existing interfaces allowing clients to use them. Both enable direct communication between clients and allow IoT deployment scenarios in which upstream connectivity is limited and local communication preferred.

\item a security analysis of our proposal, which shows that our approach
still complies with LwM2M security requirements. Our analysis considers remote
and local attackers separately and covers four common threats.

\item an empirical performance analysis conducted on real hardware and in different deployment scenarios. Our proposal outperforms a server-centric solution in terms of delay (90\%) and goodput (8$\times$).

\item open-source implementations of LwM2M client-to-client communication, which we make publicly available.
\end{enumerate}

Our extensions are carefully designed such that they reuse existing protocols defined by the LwM2M core specification.
This has two advantages.
First, our approach seamlessly integrates into the LwM2M ecosystem and, second, it allows for re-utilizing operational knowledge~\cite{gklpf-inpmm-21} and code, which is particularly important when deploying constrained devices.

The remainder of the paper is organized as follows.
\autoref{sec-lwm2m-protocol} provides the necessary background about LwM2M.
\autoref{sec-direct-lwm2m-client-communication} and \autoref{sec-third-party-auth} introduce our proposal for C2C communication and third party authorization, respectively.
\autoref{sec-security-analysis} provides a comprehensive security analysis of our proposed extensions.
Our experiments conducted on off-the-shelf IoT hardware are discussed in \autoref{sec:evaluation}, together with results revealing the advantages of C2C communication.
We present related work in \autoref{sec:related-work} and conclude with a summary and outlook in \autoref{sec:conclusion}.

\section{Background on LwM2M}\label{sec-lwm2m-protocol}

%
%

LwM2M \cite{oma-lwm2m-core-12} is a device management and service provision protocol 
that provides bootstrapping, access control, semantic data interoperability, and software update features.
Clients run on constrained devices and register themselves to one or multiple \server{s}. Machine-to-machine applications, which usually run in the cloud, interact with clients via the servers.
Server information and credentials are either pre-provisioned on a client, or bootstrapped by a
dedicated \bserver{}.

Operation semantics and parameters are first defined generically and then mapped onto the lower layer.
LwM2M supports three transport bindings: CoAP \cite{rfc-7252} (over UDP, TCP, SMS, and other Non-IP transports), HTTP, and MQTT.
Interoperability is achieved by \one~a uniform resource model and \two a RESTful interaction model \cite{ft-pdmwa-00}.
Objects are the building blocks of the resource model and specify how \client{s} group their hosted resources.
Occurrences of these groups, called object instances, contain the resources that servers access.
Multiple instances of a given object can exist on a client, each with different content but the same data structure.
Servers interact with client resources via interfaces that define operations, most of which follow a request-response scheme.

\paragraph{Overhead}
In spite of the  features provided, LwM2M adds only relatively little overhead compared to CoAP-only applications.
When analysing the processing time, our measurements reveal only $3.4\%$ of the total $\approx2570\,\mu s$  required to compute a LwM2M Read operation (\ie a GET CoAPS request).
In turn, the radio driver and the DTLS layer appeared as the dominant consumers, with $48.1\%$ and $22.4\%$ of the time respectively.
When looking at the memory footprint (see \autoref{sec:fw-size}), LwM2M represents less than $20\%$ and $33\%$ of total ROM and RAM requirements.

\paragraph{Security}
In multi-server scenarios, access control to client resources is required.
Each object instance hosted by the client has a corresponding access control object instance, that indicates the server access rights on it.
These are organized in access control lists (ACLs), where each element of the list reflects one particular server.
A single access control owner server manages the access rights and can modify policies for other servers dynamically.


LwM2M security requirements dictate clients and servers to authenticate each other, communication must be encrypted, and message integrity needs to be protected.
The protocol specifies different ways of securing communications (TLS/DTLS \cite{rfc-6347} and OSCORE \cite{rfc-8613}), for which clients need information such as URIs to uniquely identify servers, credentials (pre-shared keys, raw public keys, or certificates) and configurations (\eg security mode, ciphersuite).
Two objects organize this information: the server and the security objects, which together reflect a server account.
After establishing secure communication, clients register to servers with a unique endpoint name. In contrast to the URI, the endpoint name is independent from the transport binding.

\begin{table}[t]
    \small
    \setlength{\tabcolsep}{2pt}
    \begin{center}
        \caption{Overview of features that are required/provided~(\faCircle), partly required/provided ~(\faAdjust), or not required/provided~(\faCircleO) in different IoT scenarios/paradigms.}
        \begin{tabularx}{0.48\textwidth} {
            @{}
            >{\raggedright\arraybackslash}X
            >{\centering\arraybackslash}m
            >{\centering\arraybackslash}m
            >{\centering\arraybackslash}m
            >{\centering\arraybackslash}m
            >{\centering\arraybackslash}m
            >{\centering\arraybackslash}m}
            \toprule
            \begin{minipage}{2cm}\textbf{Scenario /\newline Paradigm}\end{minipage}
            & \begin{Turn}\textbf{Low latency}\end{Turn}
            & \begin{Turn}\textbf{High bandwidth}\end{Turn}
            & \begin{Turn}\textbf{Steady connection}\end{Turn}
            & \begin{Turn}\textbf{Sen\-si\-tive data}\end{Turn}
            & \begin{Turn}\textbf{Long range}\end{Turn}
            & \begin{Turn}\textbf{Local actuation}\end{Turn} \\
            \midrule
            Smart metering & \faCircleO & \faCircleO & \faCircleO & \faAdjust & \faAdjust & \faCircleO \\
            \dhline
            Smart farming & \faCircleO & \faCircleO & \faCircleO & \faCircleO & \faCircle & \faCircle \\
            \dhline
            Disaster first response & \faCircleO & \faCircleO & \faCircleO & \faCircleO & \faCircle & \faCircle \\
            \dhline
            Smart home & \faCircleO & \faAdjust & \faCircle & \faCircle & \faCircleO & \faCircle \\
            \dhline
            Smart transportation & \faAdjust & \faAdjust & \faCircleO & \faAdjust & \faCircleO & \faCircle \\
            \dhline
            Industrial emergency & \faCircle & \faCircle & \faAdjust & \faCircleO & \faCircleO & \faCircle \\
            \dhline
            Control systems & \faCircle & \faCircle & \faCircle & \faAdjust & \faCircleO & \faCircle \\
            \midrule
            server-centric & \faCircleO & \faAdjust & \faCircleO & \faAdjust & \faCircle & \faCircleO \\
            \dhline
            client-to-client & \faCircle & \faCircle & \faCircle & \faCircle & \faAdjust & \faCircle \\
            \bottomrule
        \end{tabularx}
        \label{table:scenarios}
    \end{center}
\end{table}

\paragraph{Shortcomings}
Despite the wide deployment of LwM2M, its server-centric paradigm presents shortcomings in certain types of use cases.
\autoref{table:scenarios} shows typical IoT scenarios, together with usually required features.
Applications such as smart agriculture and the tracking of---and interaction with---livestock present deployments at remote locations.
On the one hand, they require long-range communication (\eg to report animal vitals), on the other hand, animals need to interact with local devices (\eg gate control, food dispensing).
LoRaWAN appears as a popular long-range technology choice for this type of applications, but due to its long on-air times it applies strict duty cycles.
This quota is easily exhausted by deployments which involve control systems, as all information flows through servers even when---ultimately in many cases---a neighbour node is the recipient.

Similarly, industrial deployments involving closed-loop control systems have low latency requirements (10 -- 100 ms delays ~\cite{isa-wsiap-11}).
Such systems cannot afford a server-centric information flow due to its additional delays.
Instead, these scenarios would benefit from distributed applications based on direct local communication between \client{s}, which reduces latencies, while still being monitored by central servers.
Typically, the distributed logic is installed on the nodes after a resource discovery or a commissioning process.
Instead of a central application, nodes follow business rules (\eg "whenever the light switch of room A is pressed, notify light bulb group 2").
As they only require knowledge of a subset of the whole application, this paradigm is scalable when adding new devices.
Changes in the logic are usually performed by management tools from the central servers.

Another example are lossy IoT networks present in smart transportation containers. Given their mobile nature, they have intermittent Internet access.
The constant need for communication with the managing server requires to be permanently online.
A similar situation is faced by disaster first-response devices, which usually have to build ad-hoc delay-tolerant networks.
In these environments a connection to a central server is sporadically available.
As an alternative, the deployment of autonomous devices which can communicate with one another would allow keeping local functionalities working even if upstream connectivity is lost.

Even in scenarios with steady connectivity and high bandwidth, a central cloud involvement may raise privacy issues.
Constantly utilizing central servers to store and analyse sensor values and to control domestic appliances can reveal usage patterns and disclose personal information.
Vendors could leverage LwM2M, and install devices that interact within the household following user-installed policies.

Summarizing, LwM2M acts as a semantics-unifying layer that enables machine-to-machine applications on the cloud not only to manage IoT devices, but also to implement business logic in a vendor-independent fashion (\eg reading sensed data and activating actuators in consequence), while adding relatively small overheads.
In parallel, there is a clear need for direct node-to-node communication over a variety IoT deployments.
Consequently, we propose to extend LwM2M to allow clients to operate on each other resources, thus maintaining the benefit of its vendor interoperability and service enablement features.

\section{Direct~LwM2M~Client~Communication}\label{sec-direct-lwm2m-client-communication}




We now want to derive generic requirements for secure C2C communication in the common use cases that were analysed in the previous  \autoref{sec-lwm2m-protocol}.
As data flowing between nodes is mostly sensitive, devices are expected to establish a secure communication channel (\ie providing confidentiality, integrity, and replay protection to the messages), and to authenticate each other prior to any data exchange.
It is also desirable that resource owners are able to establish access policies to the resources (which may dynamically change at runtime), thus, some access control mechanism should be in place.
Additionally, to cope with changing deployments (\eg new appliances added to smart homes), flexibility is desired. This implies that nodes need to make use of trusted and authenticated services to securely discover resources of interest, and to obtain the required credentials and rights to access them.

%

With these requirements in mind, we enable C2C communication in the case of LwM2M, by re-utilizing existing interfaces defined in the core specification, namely
\one the device management and service enablement interface, and
\two the information reporting interface.
To avoid ambiguity when referring to \client{s} utilizing the interfaces, we define:
\one \emph{\hclient{s}} host resources on which operations are performed, and
\two \emph{\rclient{s}} request the operations on said resources.
It is worth noting, however, that nodes will likely play both roles throughout their lifetimes.
Using the interfaces requires a secure communication channel, hence, clients need to establish secure transports among each other and to have adequate access rights.
For this, we introduce a new \emph{LwM2M client account} to organize the information clients need about each other,
including security credentials, URIs and connection configurations.
Similarly to LwM2M server accounts, \client{s} hold one account per client with which they communicate.
Three newly introduced LwM2M objects organize communication and access:
\one the {\emph{client object}}, \two the {\emph{client security object} -- a \emph{LwM2M client account} consists of instances of these two objects, and
\three a \emph{client access control object} to determine which operations a \rclient{} is allowed to perform.

\subsection{Client-to-client Objects} \label{sec-client-objects}

To ease code re-utilization and lower the implementation overhead, our newly introduced objects share many resources with existing objects, used to establish client-server communication.


\paragraph{Client Object}
An instance of this object holds parameters related to the communication with other clients, including the
client ID (an internal reference), the client endpoint name, the account lifetime, default values for observation periods, and the communication binding.
To mitigate a potential elevation of privilege when access revocation messages sent by the server do not correctly arrive to the \hclient{}, the lifetime parameter in this object determines for how long a \rclient{} account is valid.
After expiration, the \hclient{} disables it, closes existing connections to the \rclient{} and ignores subsequent operation attempts.
Hence, \rclient{s} access needs periodic refresh, unless disabled by configuring the lifetime to 0.

\paragraph{Client Security Object}
An instance of this object holds the URI of a specific client, security configurations, and the DTLS credentials or a reference to an object holding OSCORE credentials, depending on which secure transport is used.
Resources of this object exhibit the exact same identifiers and semantics found in the standard LwM2M security object.
It is worth noting, that only servers are allowed to operate on client security object instances, to install and modify credentials in a dynamic fashion during the device lifetime.
This allows for additional flexibility in contrast to deployments with static configurations.
These object instances can be created and modified by servers through the device management and service enablement interface, or bootstrapped by a \bserver{}.

\paragraph{Client Access Control Object}
An instance of this object holds the actual access rights which allows \hclient{s} to keep track of the permitted operations to \rclient{s}.
Each instance is associated to a particular instance of any other object hosted by the client, and indicates an \emph{access control owner} that is the responsible server to manage access rights for this object instance.
An instance contains an ACL that specifies which operations each \rclient{} is allowed to perform on the associated instance. This is indicated using flags (\eg read, write).
For C2C access, we add an explicit `discover' access flag that controls whether a \rclient{} can explore resource attributes of an object, increasing the control granularity. This is in contrast to regular server based access, where it is always allowed to discover available objects.
Only servers can modify \rclient{s} access rights.

\begin{figure}[t]
    \includegraphics[width=0.5\textwidth]{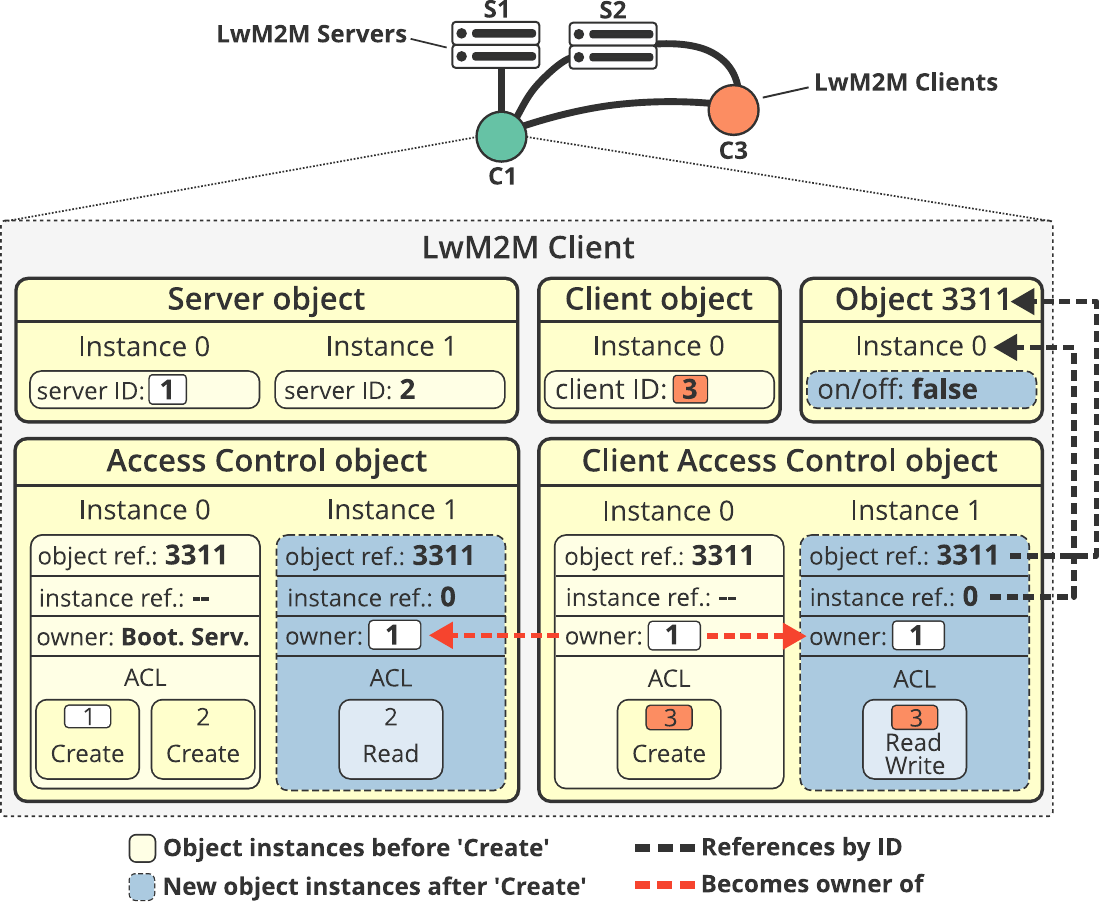}
    \caption{Access control objects in a \hclient{}, before and after a \rclient{} instantiates the light control object (ID 3311).}
    \label{fig:access-objects}
\end{figure}

\subsection{Extended Interfaces and Access Control}

Requesting clients can perform all operations defined by the device management and service enablement interface and the information reporting interface, provided they have the required access rights.
They use these interfaces to access resources in \hclient{s}, via operations like `read', `write' and `create'.
Resources may be accessible to multiple \rclient{s}, and concurrency should be handled the same way as for multiple-server access.
As per the LwM2M specification, atomicity is required when performing a `Write-Composite' operation.

All access control rules that apply to servers also do to clients.
This means that for each \rclient{} explicitly authorized to perform an operation on a resource, a corresponding ACL should be instantiated on the \hclient{}, otherwise the default access is granted.
The assignment of access control owners after a `create' operation, however, differs for C2C operations.
Whenever an object is instantiated via a `create' operation, a \hclient{} additionally creates new instances of the access control and client access control objects, to track server and client access rights respectively, for the new object instance.
In contrast to regular server operation, a \rclient{} that creates a new object instance does not become its access control owner, instead, the owner is the server indicated in the client access control object instance which authorized the `create' operation.

\begin{figure*}[t]
    \centering
    \includegraphics[width=\textwidth]{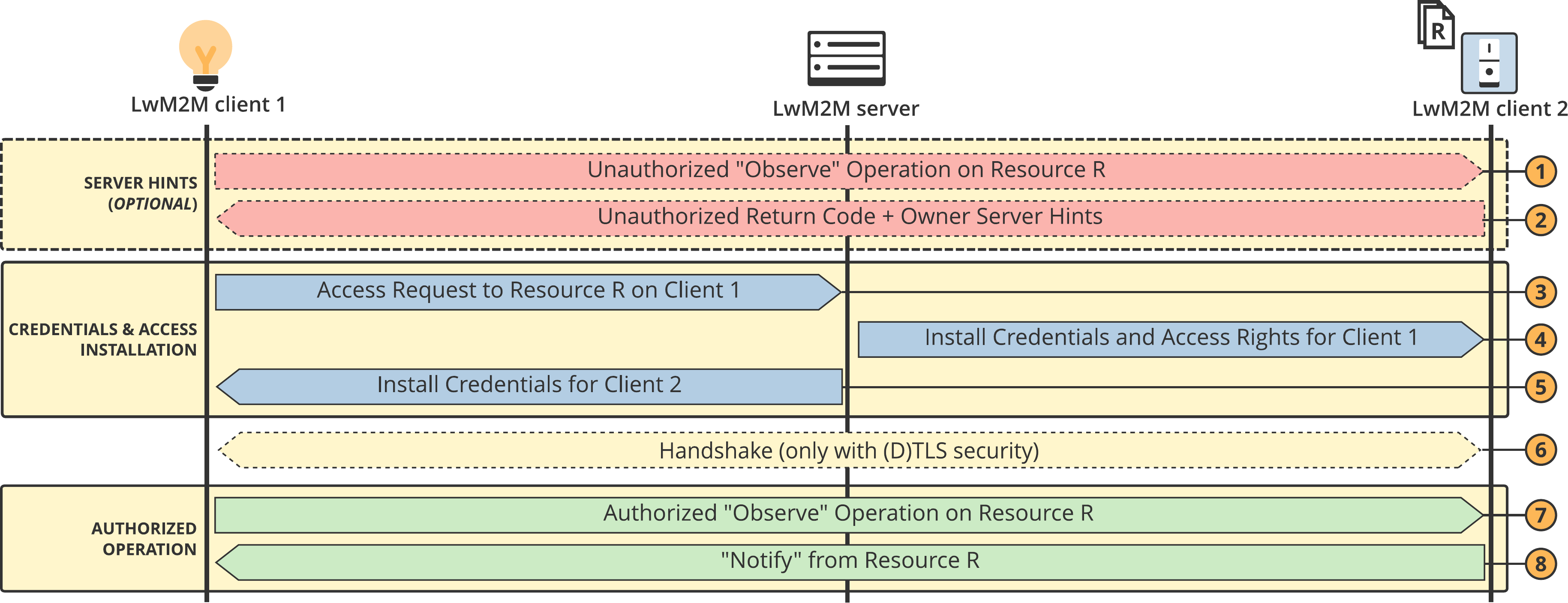}
    \caption{Request and installation of LwM2M client access rights and authorized C2C communication.}
    \label{fig:access-req-flow}
\end{figure*}

Figure \ref{fig:access-objects} illustrates a subset of the object instances on a \hclient{} $C_1$, before (yellow boxes) and after (blue boxes) a \rclient{} $C_3$ performs a `create' operation on the light control object (ID 3311).
$C_1$ hosts two server accounts ($S_1$ and $S_2$ with IDs 1 and 2), and one client account for $C_3$.
Instance 0 of the access control object holds access rights for the light control object 3311, and the ACL contains `create' access rights for both servers, whereas the \bserver{} is the access control owner.
Instance 0 of the client access control object holds `create' access rights for $C_3$, pointing to $S_1$ as the access control owner.
In both cases, as the `create' operation is performed on an object and not on a particular object instance, only the \emph{object reference} resource is set, and not the \emph{instance reference}.
When $C_3$ creates a new instance of the light control object on $C_1$, $C_1$ also creates locally new instances of both access control objects. 
Instance 1 of the access control object indicates that $S_1$ (\ie the server which allowed the instantiation) is the access control owner for the new instance of the light control object, and it grants `read' access to $S_2$.
In turn, instance 1 of the client access control object holds $S_1$ as access control owner respectively, and provides $C_3$ with `read' and `write' access rights.

\section{Third Party Authorization of LwM2M Clients} \label{sec-third-party-auth}



Clients are authorized by \server{s} that handle access rights and credential distribution. \server{s} are considered trusted third parties to the clients, as the LwM2M specification requires mutual authentication.
We introduce two new mechanisms to enable third party authorization of LwM2M clients.
\emph{Owner server hints} allow clients to discover the responsible server which can grant access to a resource.
The \emph{access request interface} is utilized by clients to request specific access rights and credentials to this responsible server.
If the server accepts the request, it distributes the access rights and credentials
through the regular device management interface.
\autoref{fig:access-req-flow} presents the complete sequence diagram of an unauthorized \rclient{} that requests access rights of a resource on a \hclient{}, utilizing the server as a third party.
The server distributes credentials, to establish a secure C2C communication, and access rights, to authorize subsequent operations among clients.
The credentials and access rights distribution is usually performed infrequently, this is, before the initial C2C interaction, but its frequency ultimately depends on the application requirements and security policies.

\subsection{Owner Server Hints}\label{subsec-owner-hints}

In multi-server LwM2M deployments \rclient{s} need to find out which server is in charge of processing a resource access request for a given node.
As LwM2M makes no assumptions as to whether servers belong to the same organization (\ie they may not communicate with one another), this knowledge must be dynamically acquired by the nodes.
To achieve this functionality in our LwM2M extension, the \hclient{} provides server information contained in the owner server hints.
It is worth noting, that this discovery mechanism can be avoided in case there is out-of-band information that a deployment has a single server.

In a first step, the \rclient{} sends an \emph{unauthorized resource request} to the \hclient{} and attempts to perform an operation on a specific resource \circledOrange{1}.
This is commonly done on unsecured transport. We further analyse the security implications in \autoref{subsec-threat-model}.
The initial contact information is assumed to be learnt by a discovery mechanism, such as a resource directory \cite{draft-ietf-core-resource-directory-26}.

A \hclient{} that receives a request on an unsecured channel from the unknown \rclient{} rejects it. It responds with an unauthorized status code, and includes the owner server hints which contain the URI of the server that owns the resource, responsible to grant access \circledOrange{2}. The response optionally contains multiple server URIs that point to separate credential- and access management servers.
To prevent disclosing information about whether a resource is hosted by the client, returned codes are kept generic and at least one default owner \server{} is included in the response, until enough trust exists with the \rclient{} and a secured transport is established.

A \rclient{} must verify that the received server URI represents a known and trusted server with which a secure communication has already been established (\ie on registration). The existing trust relation to the server is essential, since unsecured server hints could have been tampered by an attacker.



\subsection{Access Request Interface}\label{subsec-access-request-iface}


We introduce a new interface that allows clients to request servers for access rights and credentials.
The interface consists of a single operation initiated by the client: the \emph{access request}.
A client includes the intended access rights, the endpoint name of the \hclient{} it attempts to access, and a flag to indicate the need for credentials to establish secure communication \circledOrange{3}.
Utilizing \emph{endpoint client names}~\cite[Sect. 7.3.1]{oma-lwm2m-core-12} abstracts from the underlying protocol and allows to identify clients across LwM2M transport bindings. As a counterexample, URIs reveal different structures across transport bindings which complicates interoperability in heterogeneous deployments. It is noteworthy that this requires unique endpoint names for clients that participate in a network.

After reception of the access request, the server verifies if the client is entitled to obtain the requested access rights. This decision depends on the application-logic. Commonly servers follow pre-installed access policies or query the resource owner.
On acceptance, the server generates credentials and creates client accounts on both clients. Thereafter, it modifies the client access control object of the \hclient{} to enable the \rclient{} to access to the required resource \circledOrange{4}\,\&\,\circledOrange{5}.
Once clients are in possession of the credentials, if they are using (D)TLS-based security they perform the handshake \circledOrange{6}, and if they use OSCORE the derive locally the corresponding security contexts.
Finally, authorized client-to-client interactions can be performed \circledOrange{7}\,\&\,\circledOrange{8}.

We present a mapping of the access request interface onto CoAP transport, however, our approach naturally extends to other LwM2M transport protocols.
A \rclient{} performs an access request operation sending a \texttt{POST} request on the path \emph{/ac} to the \server{}.
The operation has two parameters passed as URI query strings:
\one \emph{ep} is mandatory and holds the endpoint client name of the \hclient{}.
\two \emph{c} is optional and indicates if the \rclient{} requires credentials to be installed in order to initiate secure communications with the \hclient{}.

The requested access rights are included in the payload of the request, encoded in LwM2M CBOR \cite{rfc-7049} format.
A message can contain multiple access requests to multiple objects and instances.

\section{Security Analysis}\label{sec-security-analysis}

\begin{table*}[]
    \small
    \setlength{\tabcolsep}{2pt}
    \begin{center}
        \caption{Threat model of the LwM2M client-to-client communication and third party authorization extensions.}
        \begin{tabularx}{\textwidth}{cXccXcX}
            \toprule
            \textbf{No.}&\makecell[c]{\textbf{Threat description}} & \makecell[c]{\textbf{Asset}\\(\S\ref{subsec-assets})} &  \makecell[c]{\textbf{Adversary}\\(\S\ref{subsec-attacker-model})} &  \makecell[c]{\textbf{Surface}\\(\S\ref{subsec-attack-surface})} & \makecell[c]{\textbf{CIAA}\\(\S\ref{subsec-assets})} & \makecell[c]{\textbf{Mitigation}} \\
            \midrule
            T0
            & \textbf{Information Disclosure}: Observing unprotected operations attackers can learn which resources may be hosted and which are of interest.
            & \makecell[t]{App. config. /\\Client\\resources.}
            & \makecell[t]{Local}
            & \makecell[t]{Unauthorized request /\\Server hints.}
            & CO
            & Sensitive content should be avoided on unauthorized requests. \\
            \midrule
            T1
            & \textbf{DoS}: Open DTLS port on a client is used for message amplification to perform a denial of service attack.
            & \makecell[t]{Operational\\resources.}
            & \makecell[t]{Remote}
            & \makecell[t]{Open client DTLS port.}
            & AV
            & The DTLS server sends out a \texttt{HelloVerifyRequest} message during handshake. \\
            \midrule
            T2
            & \textbf{Elevation of privilege}: A server cannot revoke client access to a resource, elevating its privilege, because incoming communication is jammed.
            & \makecell[t]{Hosted\\resources.}
            & \makecell[t]{Local}
            & \makecell[t]{Extended device\\management interface}
            & CO
            & Lifetime parameter in the client security object defines maximum period of access. \\
            \midrule
            T3
            & \textbf{Tampering}: A \rclient{} receives invalid server hints, which might point to a rogue or compromised server.
            & \makecell[t]{Owner\\server URI.}
            & \makecell[t]{Local}
            & \makecell[t]{Server hints}
            & IN
            & \client{s} only consider for access requests \server{s} to which they are already registered. \\
            \bottomrule
        \end{tabularx}
        \label{table-threat-model}
    \end{center}
\end{table*}

Throughout the design of the proposed LwM2M extensions we followed an iterative threat-driven approach,
by performing a methodical four-steps analysis of the security and privacy aspects of the extensions:
\one asset identification and security properties assignment (\autoref{subsec-assets}),
\two attacker model building (\autoref{subsec-attacker-model}),
\three attack surfaces analysis (\autoref{subsec-attack-surface}),
\four threat analysis, which lead to protocol improvements.
In this section, we present each step of the analysis and the outcome of the threat model, together with mitigations.



\subsection{Assets and Security Properties}\label{subsec-assets}

In this step we enumerate the resources with security properties that should be preserved, and that might be targeted by attackers.
We use the well known CIA(A) \cite{nist-fips199-04} descriptors as the space of security properties:
Confidentiality (CO), Integrity (IN), Availability (AV), and Authenticity (AU).


\paragraph{Application configuration (CO, IN)} Considers the node behaviours and their relations with other clients and servers, \eg the interest of client $C_3$ in resource R hosted by client $C_1$.

\paragraph{Owner server URI (IN)} Identity of a resource owner, \eg the server that assigns access rights. It is worth noting, however, that confidentiality of the owner server URI is not expected, as it is usually sent over un-protected communication channels between clients.

\paragraph{Client access rights (CO, IN)} Access grants of \rclient{s} to resources on \hclient{s}, \eg permissions that the owner server grants a client $C_3$ on resources hosted by client $C_1$.

\paragraph{Client credentials (CO, IN, AU)} Key material used for secure communication, \eg the pre-shared key of a client.

\paragraph{Hosted client resources (CO, IN)} LwM2M resources on a \hclient{}, \eg the status of a light control object.

\paragraph{Device operational resources (CO, IN, AV, AU)} The preservation of networking-, computational-, battery-, and memory resources, \ie a device that hosts/operates on a resource R remains in operation.

In addition to preserving the aforementioned asset properties, we must consider the LwM2M security requirements defined in \cite[Sect. 5.1]{oma-lwm2m-transport-12}, which apply across all transport bindings.
They state that
\one messages must be replay protected,
\two requests and responses must be bound,
\three freshness must be verifiable for certain operations,
\four secure fragmentation must be supported,
\five data from clients and servers must be encrypted and integrity protected and
\six clients and servers must be authenticated prior to data exchange.

\subsection{Attacker Model}\label{subsec-attacker-model}

Our attacker model assumes that adversaries are not in possession of valid credentials required for mutual authentication with the server or clients.
We identify two attacker groups:

\paragraphS{Remote attackers} access nodes remotely through the network.
They may be capable of eavesdropping messages transmitted between clients and servers but have no access to the local client network.
These attackers try to learn internals and use this information to compromise a device under attack.
Therefore, they impersonate \hclient{s}, \rclient{s} or \server{s} by sending malicious messages via the LwM2M interfaces.
Remote attackers usually leverage protocol or software vulnerabilities to manipulate sensitive processing tasks.

\paragraphS{Local attackers} have direct access to the local network.
Additionally to the capabilities of a remote attacker, local ones may intercept, modify and replay messages among clients on the local area network (usually wirelessly).
Attackers who apply advanced hardware access techniques to manipulate secret information directly from the silicon are not in the scope of this analysis.

\paragraph{On peer-to-peer attacks}
 LwM2M deployments are commonly composed of heterogeneous embedded devices with specific capabilities, resources and tasks that are all part of a single administrative domain. 
Even though we analyse direct communication between \client{s}, these networks are not equivalent to traditional peer-to-peer (P2P) systems, which run on independent responsibilities.
Consequently, we consider typical P2P attacks out of the scope for this analysis.
As an example, rational attacks would not apply because nodes participating in the LwM2M deployment are naturally cooperating (\ie they expose their resources and share via the LWM2M server).

\subsection{Attack Surfaces}\label{subsec-attack-surface}

Attack surfaces are potential entry points that can be leveraged by adversaries to perform attacks against the system assets.
Here, we analyse the attack surfaces of the LwM2M extensions.

\paragraph{Open DTLS port on client} To allow secure DTLS-based transport among clients, these need to accept incoming session handshakes, similarly to LwM2M servers.
It is noteworthy, however, that OSCORE \cite{rfc-8613} as the alternative secure transport requires no handshake.

\paragraph{Extended device management interface} Not only servers, but also other clients, can access client resources.
The increased number of accessing devices, which can be potentially compromised, enlarges this attack surface compared to server-centric LwM2M deployments.

\paragraph{Unauthorized resource request} In order to learn which server owns access rights to a resource, \rclient{s} may perform unauthorized requests, which often occur over non-secured transports.

\paragraph{Server hints} In response to an unauthorized request, a \hclient{} responds with the owner server hints, which may be sent in clear text prior to establishing a secure transport.


\subsection{Threat Model}\label{subsec-threat-model}

\autoref{table-threat-model} presents a series of threats that we identify based on the former analysis of assets, adversaries, and attack surfaces.
To classify the threats we follow the STRIDE framework~\cite{kg-top-99} which defines six categories of security threats:
Spoofing identity (S),
Tampering with data (T),
Repudiation (R),
Information disclosure (I),
Denial of service (D) and
Elevation of privilege (E).

T0 describes a threat in which an attacker, who eavesdrops an unprotected unauthorized resource request, acquires information about a possible resource hosted by the \client{} and the interest of the \rclient{} on it.
An attacker learning this information may raise a privacy issue, thus, \rclient{s} should avoid sending sensitive payload on unprotected unauthorized request (\eg only perform read operations), and \hclient{s} should keep response codes generic.
In cases where particular resource URIs must not be revealed, a \rclient{} can perform an initial request to a non-sensitive resource to get the owner server hints, from which it can request initial credentials.
After the secure channel has been established, the sensitive request can be performed.

T1 is identified as a threat introduced by the extensions, because in a server-centric LwM2M deployment clients would not play the server role during a DTLS handshake, and could be configured to simply ignore them, when using DTLS-based security.
One possible mitigation is to use the \texttt{HelloVerifyRequest} message with a stateless cookie, making the usage of spoofed IP addresses for DoS attacks difficult.
Another strategy is to establish a limit for incoming requests.

T2 and T3 are mitigated by design in the proposed LwM2M extensions. T2 considers a situation in which an access right revocation message does not arrive to the \hclient{}, either because an attacker blocks it or because of the lossy nature of the networks. This results in an elevation of privileges for the \rclient{}, who keeps the access beyond the intended period.
By assigning a lifetime to the distributed credentials the impact of such an attack is reduced, at the cost of an increased traffic generated by periodic authorization requests.
T3 considers the case of invalid server hints, which are sent to a \rclient{} (\eg injected by a local attacker) and could point to a compromised or rouge \server{}. A \rclient{} should only consider known servers as valid owners to request access credentials.

Now we analyse the compliance of the extensions with the LwM2M security requirements (listed in \autoref{subsec-assets}).
As C2C communication is not considered in the specification, we give the requirements a broad scope to consider data exchanged among clients as well.
Requirements \one through \four are fulfilled by the underlying transport bindings, as we utilize the same ones as in standard LwM2M deployments for C2C communication.
Messages exchanged over the extended device management and access request interfaces are encrypted and integrity protected (requirement \five) by both OSCORE and DTLS.
Moreover, these protocols also provide mutual authentication (requirement \six) to clients when performing operations on the extended interface.
The only messages sent prior to mutual authentication, and that are not encrypted nor integrity protected, are the initial unauthorized resource request and the server hints.
We have already described the impact of this and provided mitigations to reduce the exposure through this surface. Only the URIs of the requested resource and \server{s} would be sent over unprotected transport, and no critical information would be disclosed.
In the case when a particular application cannot afford such disclosure, \client{s} can be pre-provisioned with credentials and still establish a C2C communication.
Thus, we conclude that the extensions comply with the LwM2M security requirements.

\section{Performance Evaluation}\label{sec:evaluation}
In this section, we compare our proposal, client-to-client communication and authorization, with the current client-server architecture in LwM2M.
We analyze memory consumption, transmission delays, and maximum goodput, based on experiments on real hardware.
Our experiments are guided by the use cases of edge processing and distributed application logic in single- and multi-hop deployments.

\begin{figure*}[ht]
    \centering
    \begin{subfigure}[t]{0.48\textwidth}
        \centering
            \includegraphics[width=\textwidth]{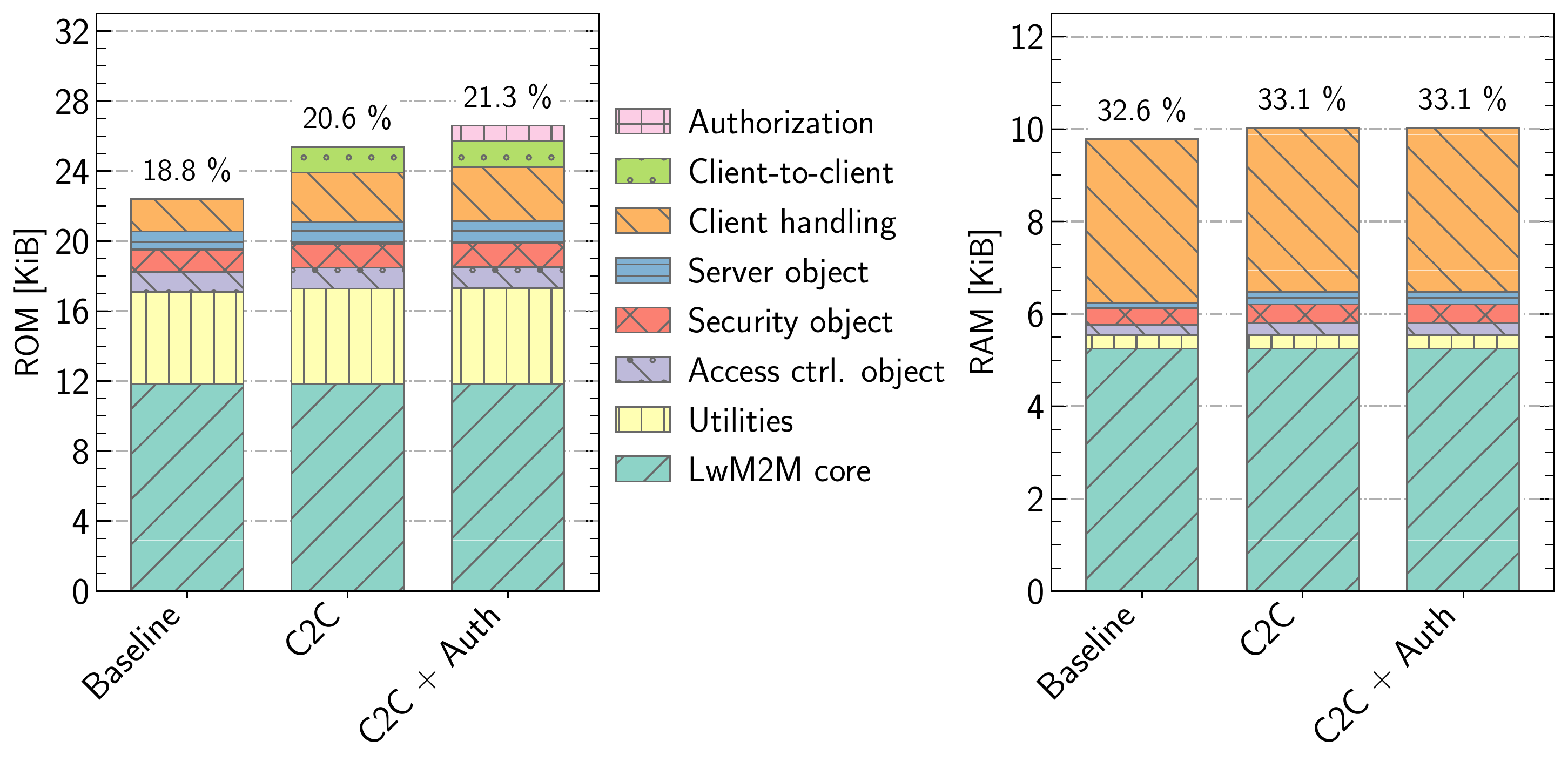}
            \caption{ROM (left) and RAM (right) requirements of LwM2M client modules in a default configuration (Baseline), with client-to-client (C2C) and third party authorization (Auth) extensions. Relative values relate to the complete firmware image size.}
            \label{fig:size-impact}
    \end{subfigure}
    \hfill
    \begin{subfigure}[t]{0.48\textwidth}
    \centering
        \includegraphics[width=\textwidth]{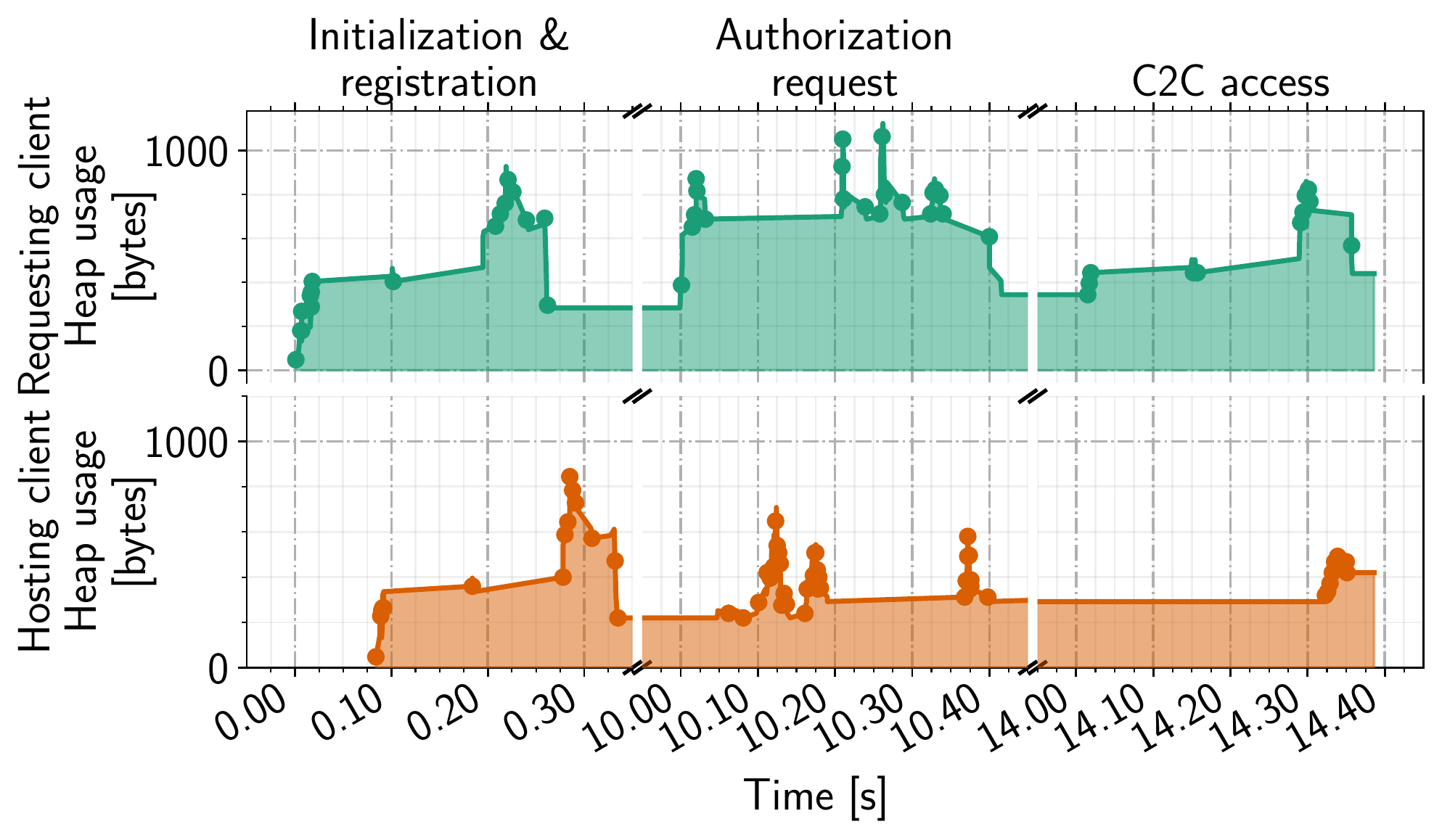}
        \caption{Heap usage of a LwM2M \rclient{} (top) and a LwM2M \hclient{} (bottom) during initialization and registration, authorization request, and C2C access.}
        \label{fig:heap-usage}
    \end{subfigure}%
    \caption{Memory requirements of LwM2M client modules (~\ref{fig:size-impact}) and heap usage by \rclient{} and \hclient{} (~\ref{fig:heap-usage}).}
\end{figure*}

\subsection{Experiment Setup}

\paragraph{Hardware and Software} We conduct our experiments by deploying real implementations on the FIT IoT-LAB testbed, using off-the-shelf class 2 IoT devices \cite{RFC-7228} that feature ARM Cortex-M3 MCUs running at 72 MHz,
with 64 KiB of RAM and 512 KiB of ROM, equipped with IEEE 802.15.4-compatible Atmel AT86RF231~transceivers.

The firmware that runs on the constrained IoT~devices is based on the operating system RIOT, version \texttt{2021.04}.
The \client{} is implemented using Wakaama version \texttt{1.0}, the heap, needed by Wakaama, with the Two-Level Segregated Fit (TLSF) allocator.
In terms of security support, we consider both OSCORE (the current main branch of uOSCORE) and DTLS (current development branch of tinyDTLS).
On the \server{} side, we use the implementation Leshan, version \texttt{1.3.1}, which runs on a Dell PowerEdge R6525 server with two AMD EPYC 7702 processors and 512 GB of~RAM.

\paragraph{Configuration and Startup} In all DTLS and OSCORE deployments, we use AES in CCM mode with a 128-bit key, an authentication tag of 8 bytes, and pre-shared keys (PSK) suites.
The testbed hardware does not feature a hardware random number generator.
To generate the initial CoAP message ID, we use the SRAM-based physically unclonable function (PUF) \cite{ksw-gpngi-21} as entropy source to seed the RIOT pseudo random number generator.

\paragraph{Deployment Scenarios} To compare the performance of C2C versus server-centric communication, we deploy multiple topologies running three scenarios:
\one server-centric,
\two C2C using DTLS security,
\three C2C using OSCORE security.
In all scenarios, a \hclient{} produces a 5-bytes data every second, which should reach a \rclient{}.
We focused our experiments on the observation of resource updates, as it is, for the typically deployed low-power sleepy devices, a more common approach than constant polling.
In scenario \one, a centralized application interacts with the \server{} via an HTTP API and observes the sensor resource, writing new values to another client upon update notifications.
In scenarios \two \& \three, the application logic is decentralized, \ie the \rclient{} observes the sensor resource in the \hclient{}, thus, receiving periodic notifications directly.

\subsection{Firmware Size}\label{sec:fw-size}

\autoref{fig:size-impact} presents memory requirements for three configurations of the LwM2M client firmware:
\one~baseline (no extensions),
\two~C2C extension enabled, and
\three~C2C + Auth extensions enabled.
Measurements are separated into ROM, which considers the code segment and variables initial values (\texttt{text} + \texttt{data} segment), and RAM, which includes (un-)initialized global variables (\texttt{bss} + \texttt{data} segment).
ROM and RAM consumption of the LwM2M core module remain unaffected across configurations and require $\approx$ 11\,KiB ROM and 5\,KiB RAM, including the RAM memory pool used by the heap allocator.
Similarly, utilities modules are fundamentally constant, however, our Auth extension adds 460 bytes of ROM for introducing CBOR encoding which would also benefit the pure Wakaama baseline implementation.

Due to the functional similarities between the newly introduced LWM2M objects and their existing counterparts, we are able to reutilize most of the code (\ie no extra C object files are compiled for the new LwM2M objects), only with slight size increments to accommodate the extra logic.
The size of security, access control, and server objects increase by $\approx$ 100~bytes, 110~bytes, and 200~bytes in ROM, and $\approx$ 40, 40 and 170 bytes in RAM, due to the additional states to handle client security, client access control, and client objects.
The client handling module is responsible for connections and \rclient{} operations, which adds 970 bytes in ROM within the C2C extension, and additional 310 bytes with the Auth extension, for additional logic of connection handling and client credential management.
The C2C and Auth modules use 1230 and 910 extra bytes of ROM, while no extra RAM in needed, since connection states are stored in the security and server objects.
Memory requirements that correspond to the OS, network stack, and drivers have a constant memory offset across configurations (not displayed in~\autoref{fig:size-impact}), however, we indicate the percentage of our LwM2M client modules to the total firmware size.
In summary, the LwM2M proportions conform $\approx$ 20\,\% of the ROM and $\approx$ 33\,\% of the RAM in comparison to the total image, which is around 125\,KiB in ROM and 31\,KiB in RAM.

We conclude that the pure overhead of our C2C extension increases total ROM image size by only $\approx$3.3\,\% and RAM by 0.9\,\%, while the Auth extension requires additional $\approx$ 5.0\,\% of ROM and no extra RAM. This is in line with our goal to maximize code re-utilization.

\begin{figure*}[t]
    \centering
    \includegraphics[width=\textwidth]{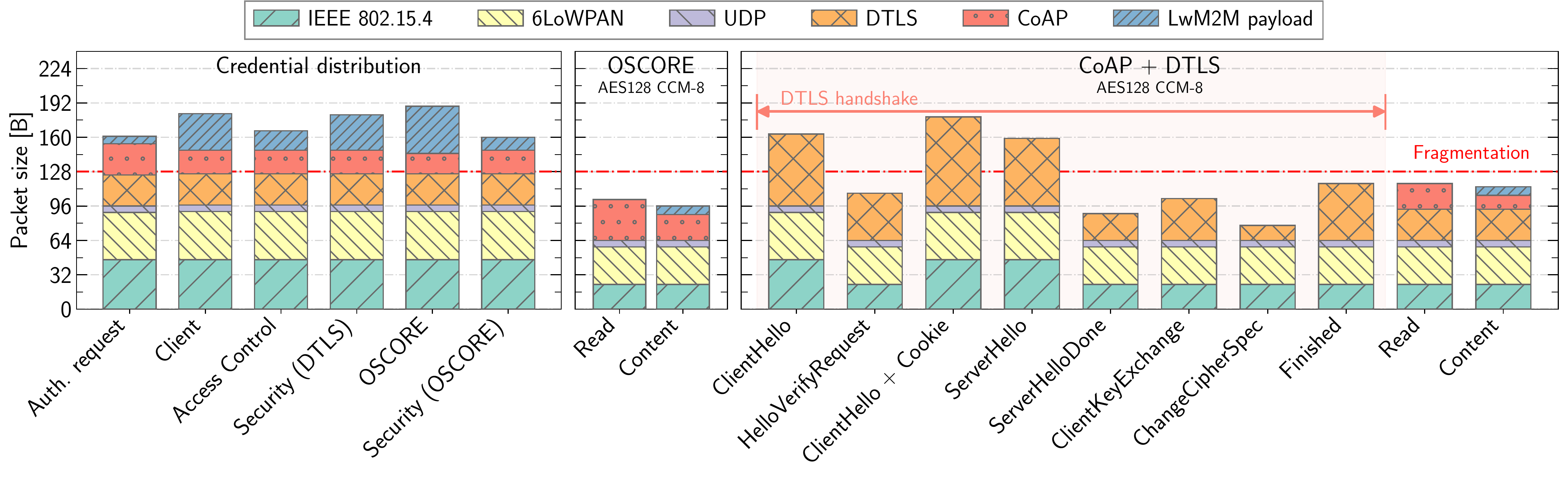}
    \caption{Packet structures of credential distribution and client-to-client read operation, using DTLS and OSCORE.}
    \label{fig:packet-structure}
\end{figure*}

\subsection{Heap Usage}\label{sec:dyn-memory}

\autoref{fig:heap-usage} illustrates heap usage on two clients, separated into three phases: Initialization \& registration, authorization request, and C2C access. The top half of the figure corresponds to a LwM2M \rclient{}, and the bottom half to a LwM2M \hclient{}. We exclude the LwM2M server memory consumption, since it does not suffer from memory constraints.
Overall, heap requirements range from $\approx$ 300--1100\,bytes on constrained nodes.

The usage pattern during initialization \& registration (00.00--00.35\,seconds) is similar for both clients, however, the \rclient{} requires additional $\approx$ 80\,bytes to allocate a structure that holds initial information about the LwM2M \hclient{}. This memory is required for every \hclient{} to which the \rclient{} connects. In contrast, the host has no prior knowledge of a \rclient{}.
At 10\,seconds, the \rclient{} initiates an authorization request to the server which requires memory for the state of a CoAP request.
A spike in the graph at 10.03 seconds corresponds to temporary memory used for the CBOR message encoding.
Between 10.10--10.38\,seconds, spikes in both graphs correspond to read and write operations on the resources, performed by the server, which installs credentials and access rights.
After the authorization request has finalized, at 10.40 seconds, the \rclient{} de-allocates its state and memory usage returns to a new baseline slightly higher than the previous one, because of the new client information that Wakaama needs to allocate internally.
The \rclient{} starts a DTLS handshake with the \hclient{} at 14.02\,seconds. An increment of $\approx$ 100\,bytes reflects an active DTLS connection, that has to be allocated and kept for every host, until the connection closes.
Finally, at 14.28\,seconds, a C2C read operation allocates $\approx$ 200\,bytes state on the \rclient{}.
Upon reception, at 14.34\,seconds, the \hclient{} allocates heap memory for the new DTLS connection ($\approx$ 100\,bytes similarly to the \rclient{} on handshake) and utilizes temporary memory to format the response message.

\subsection{Packet sizes}

Now we dissect the packets that constitute the authorization request flow and a C2C read operation, using both DTLS and OSCORE security, as shown in \autoref{fig:packet-structure}.

The maximum data unit size of the IEEE 802.15.4 2.4 GHz physical layer is 127 bytes.
Considering the sizes of the 8-bytes destination and source hardware addresses,
2-bytes frame control field, 1-byte sequence number, 2-bytes personal area network (PAN) ID, and 2-bytes frame check sequence (FCS),
the MAC header adds to 23 bytes, which allows up to 104 bytes to be transmitted by the upper layers.
A total of 41 bytes are used by the 6LoWPAN layer, as it requires 2 bytes to accommodate the IP header compression, 1 for the hop limit, 32 for both IPv6 addresses, and 6 to encode the compressed UDP header.

The authorization request is for read access on one object instance, encoded in CBOR as detailed in \autoref{subsec-access-request-iface}.
During credential distribution the authorization request, client and access control object instantiation messages are common across transports.
On the other hand, the content of the security object instantiation message only holds credentials when using CoAP over DTLS (DTLS security), as OSCORE credentials are distributed separately in its own OSCORE object.
All messages in this process trigger 6LoWPAN fragmentation as they are bigger than the physical data unit.
The object instantiations and the content messages are encoded in LwM2M TLV, because of the current support in Wakaama.
The OSCORE read and content packets are respectively 15 and 21 bytes smaller than the DTLS counterpart, due to the bigger size of the DTLS record layer compared to the OSCORE header.

\subsection{Time to resource update}

\begin{figure}[]
    \centering
    \includegraphics[width=0.5\textwidth]{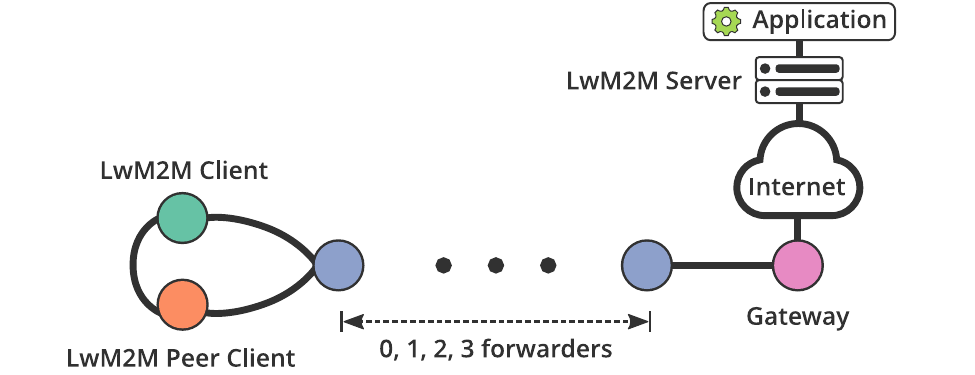}
    \caption{A variable number of forwarders to determine the impact of multihop on server-centric LwM2M deployments.}
    \label{fig:topology-variable-forwarders}
\end{figure}

\begin{figure*}[]
    \centering
    \begin{subfigure}[t]{0.48\textwidth}
    \centering
        \includegraphics[width=\linewidth]{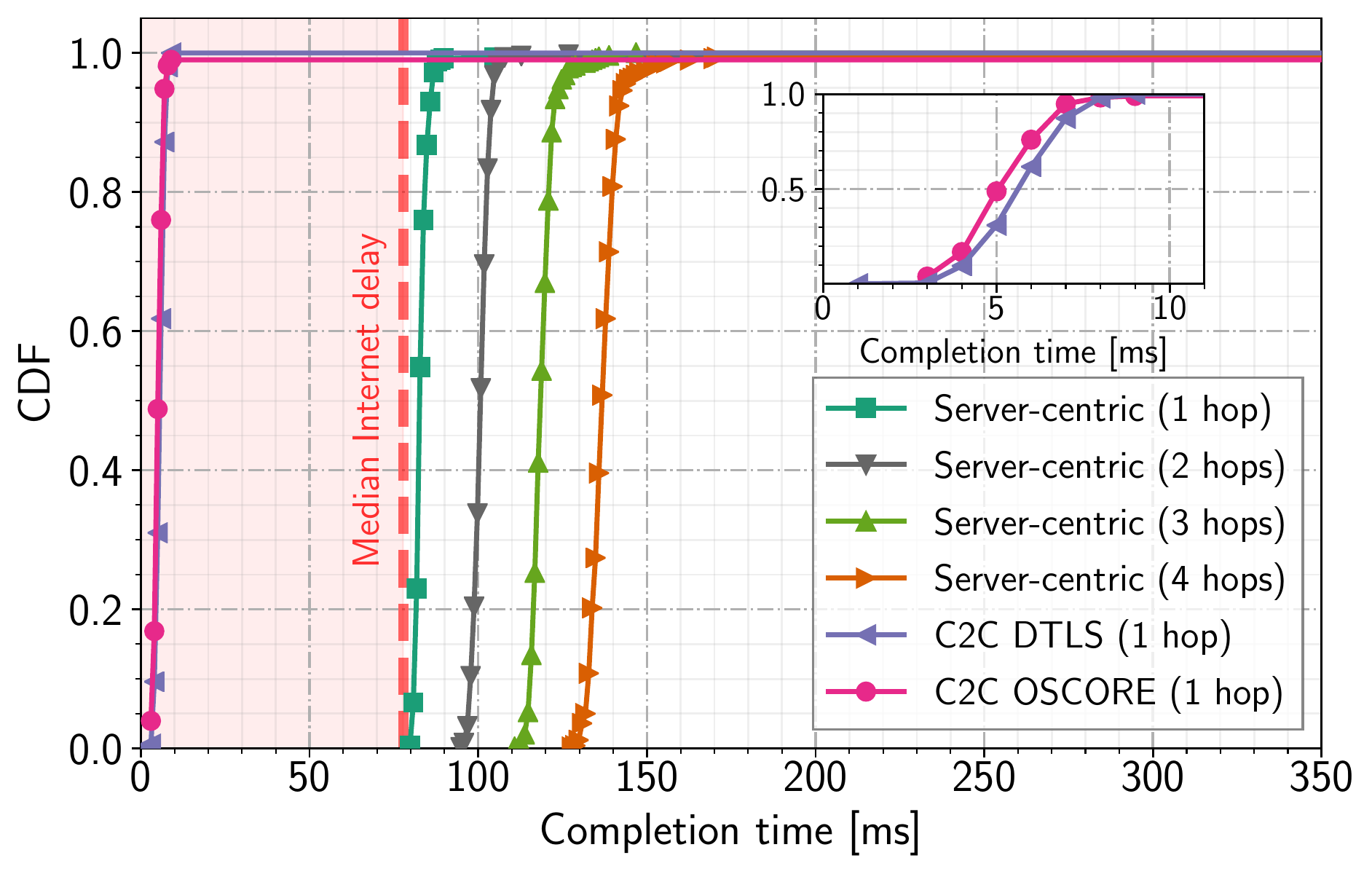}
        \caption{Notification arrival times in server-centric and C2C deployments.}
        \label{fig:notification-completion}
    \end{subfigure}%
    \hfill
    \begin{subfigure}[t]{0.48\textwidth}
    \centering
        \includegraphics[width=\linewidth]{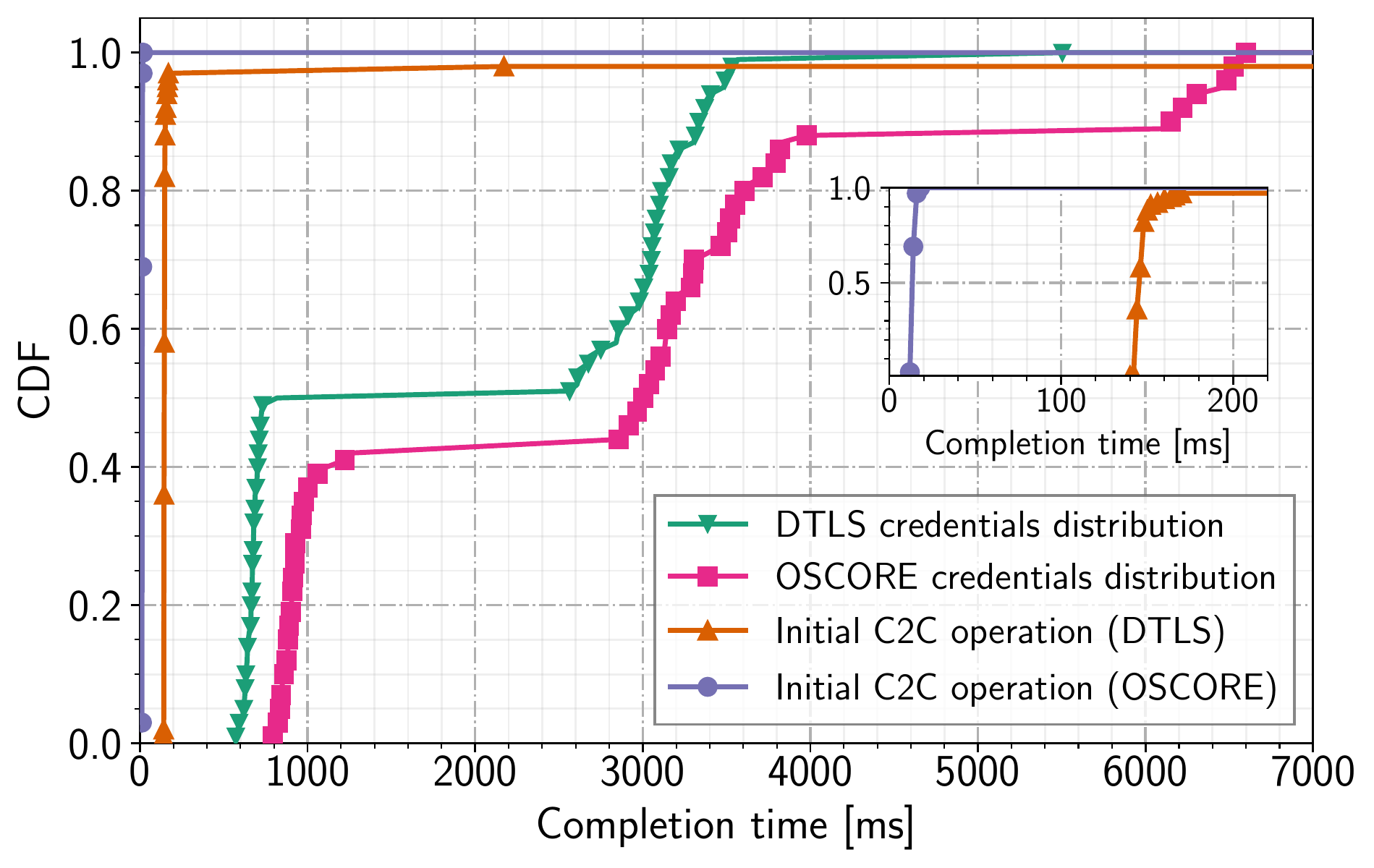}
        \caption{Authorization request and initial C2C operation completion times.}
        \label{fig:authorization-completion}
    \end{subfigure}
    \label{fig:completion-times}
    \caption{Temporal distributions of notification arrival (~\ref{fig:notification-completion}) and authorization request followed by first C2C operation (~\ref{fig:authorization-completion}).}
\end{figure*}

We analyse the times between the generation of a new resource value in a \hclient{} and its arrival at a \rclient{} on all deployment scenarios.
The topology for this experiment consists of \client{s} connected to the \server{} through a gateway and a varying number of intermediate forwarder nodes, as shown in \autoref{fig:topology-variable-forwarders}.
We vary forwarders to quantify the impact of extra hops when using the server-centric LwM2M deployment.
\autoref{fig:notification-completion} shows the results for C2C and server-centric communication for different amount of hops between the gateway and the clients.
To reduce the impact of the variable delay introduced by the Internet connection between the gateway and the \server{}, we first subtracted this time from the server-centric measurements,
and then offset them by the median Internet delay of across all experiments \mbox{($\approx$ 78 ms)}, which was measured by timestamping the packets from and to the gateway.
We observe a reduction of $\approx$ 90\% in the notification delay when using C2C communication compared to the server-centric single-hop scenario.
The extra time required in the latter scenario is explained by the overhead of the 'write' operations from the \server{} to the client and the delays of the connection between the gateway and the server.
The impact of additional hops is of $\approx$ 15 ms delay per hop when communicating through the server, which is consistent with typical 6LoWPAN times.

Now we consider a second setup where the \rclient{} first performs an authorization request to the \server{} and then a C2C read operation.
The entire credential distribution is composed of 5 operations:
\one authorization request,
instantiation of
\two client object,
\three client security object and
\four client access control object in \hclient{},
\five update of the client security object instance in \rclient{}.

\autoref{fig:authorization-completion} shows that $\approx$ 50\% of the DTLS credential distributions are completed in less than 1 second without needing CoAP retransmissions, while within 4 seconds most of the DTLS credentials distributions are successful.
We observe that distributing OSCORE credentials takes slightly longer and needs a second retransmission for $\approx$ 15\% of the cases, due to the instantiation of one more object compared to DTLS credentials (the OSCORE object).
We can observe a stair-case pattern caused by CoAP retransmissions that reflects the default configuration of a 2-seconds ACK timeout \cite{rfc-7252}.
Once credentials are distributed, the initial C2C read operation using OSCORE takes $\approx$ 10 ms, while the initial DTLS handshake raises this time to between 140 and 160 ms.

Next, we look at the effective goodput achieved across deployment scenarios, summarized in \autoref{fig:throughput}.
For this experiment, a \hclient{} sends 5.000 notifications at varying intervals, configuring the radios at 250 Kbit/s and 2000 Kbit/s (minimum and maximum available values respectively).
The resulting goodput measurements are depicted using box plots, next to the theoretical optimum (dashed lines).
For each interval the rate of successfully delivered notifications is plotted as well.
We can observe an almost optimal behaviour in both C2C scenarios, with a steady delivery rate close to 100\%, which only starts to degrade when approaching 10 ms intervals. This is in line with the times shown in \autoref{fig:notification-completion}.
On the other hand, the server-centric scenario reveals a maximum LwM2M payload goodput of $\approx$ 50 B/s, and a degradation of the delivery rate for intervals bellow 100 ms, which is in concordance with the notification arrival times we observed before. When utilizing a higher radio data rate we observe a slight improvement in the delivery rates and less dispersion in the goodput values.
We attribute this to a lower probability of packet collision.

Finally, to simulate a more realistic and less controlled topology, we construct six topologies of 20 randomly selected nodes each.
The constraints for the selection algorithm are a minimum distance of 2.2 m and a maximum of 6.6 m between nodes, which has resulted in a sufficiently reliable communication for our measurements.
For each topology, the \hclient{} and the \rclient{} are also randomly chosen.
\autoref{fig:topologies-notification} depicts the randomly built topologies and the number of hops between the nodes of interest.
We measure the notification completion times for the three previously-described deployment scenarios, and observe that C2C performs between 60\% and 90\% faster than server-centric communication.

\begin{figure*}[]
    \centering
    \includegraphics[width=\textwidth]{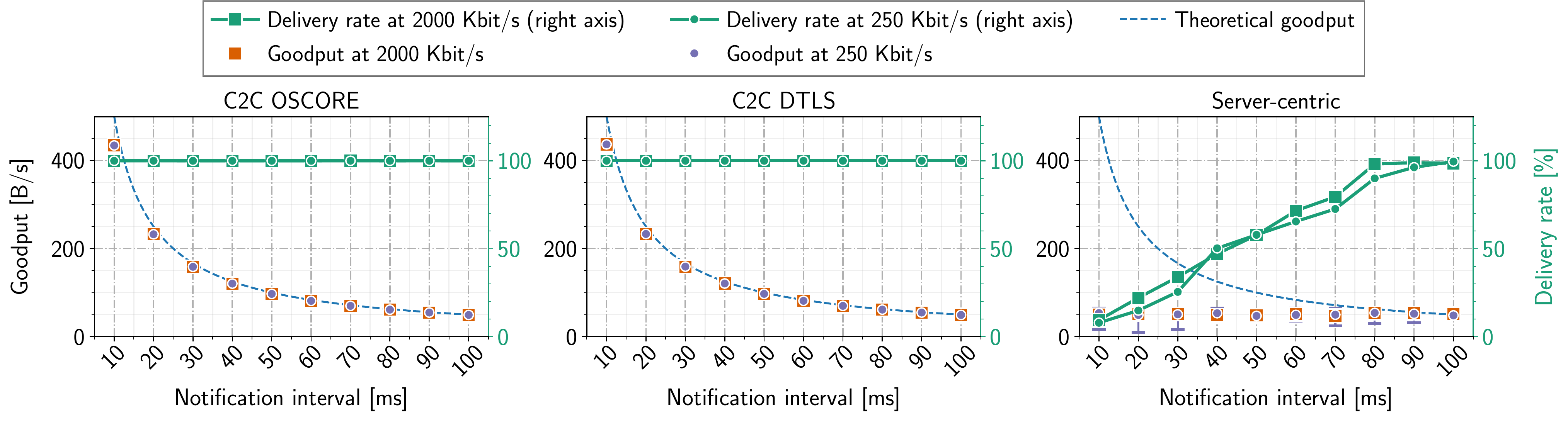}
    \caption{Goodput of client-to-client (C2C) and server-centric deployments using different notification intervals.}
    \label{fig:throughput}
\end{figure*}

\subsection{Energy consumption}

For the topologies depicted in \autoref{fig:topology-variable-forwarders} we now measure the energy consumption while sending notifications to a \rclient{}.
\autoref{fig:energy} shows the energy consumption aggregated through all nodes but the gateway, which was not considered energy-constrained due to its wired connection.
We observe that the main impact in energy consumption occurs when increasing the number of intervening nodes, and that C2C deployments, at $\approx$ 42.9 J, pay no energy overhead for the features, nor for the simultaneous utilization of DTLS and OSCORE.
Moreover, the right side of \autoref{fig:energy} shows that forwarder nodes require $\approx$ 12\% more power than hosts and requesters.
We can conclude that C2C communication helps relaxing the overall energy requirements in LwM2M deployments by reducing the amount of intermediate forwarder nodes.

\begin{figure}[]
    \centering
    \includegraphics[width=0.5\textwidth]{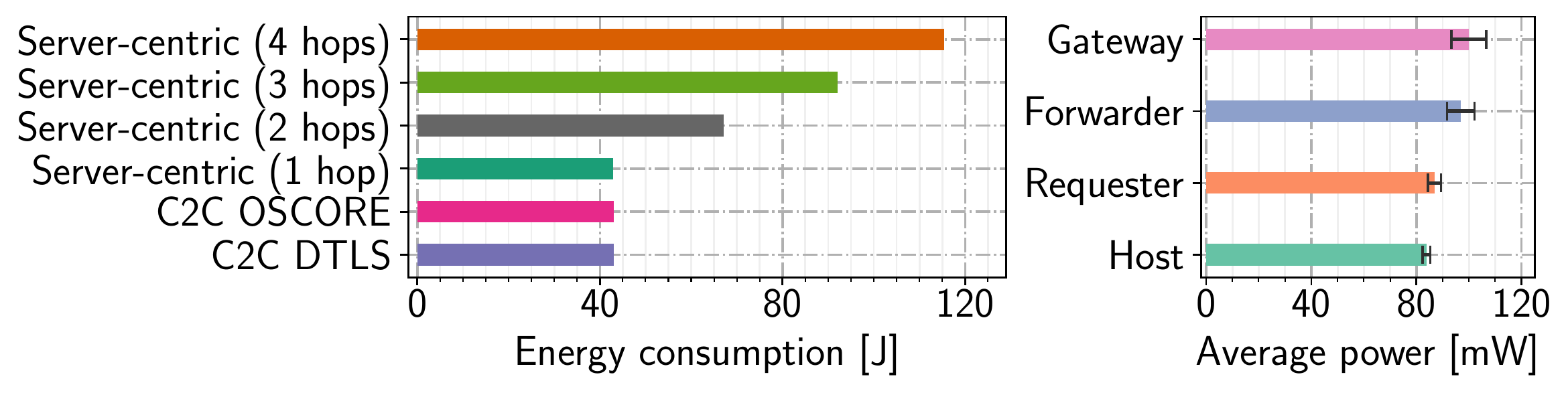}
    \caption{Total energy consumption for different deployments (left), and the average with the standard deviation of power requirements for different node types (right).}
    \label{fig:energy}
\end{figure}

\begin{figure*}[t]
    \centering
    \begin{subfigure}{0.3\textwidth}
        \centering
        \includegraphics[width=\textwidth, height=4cm]{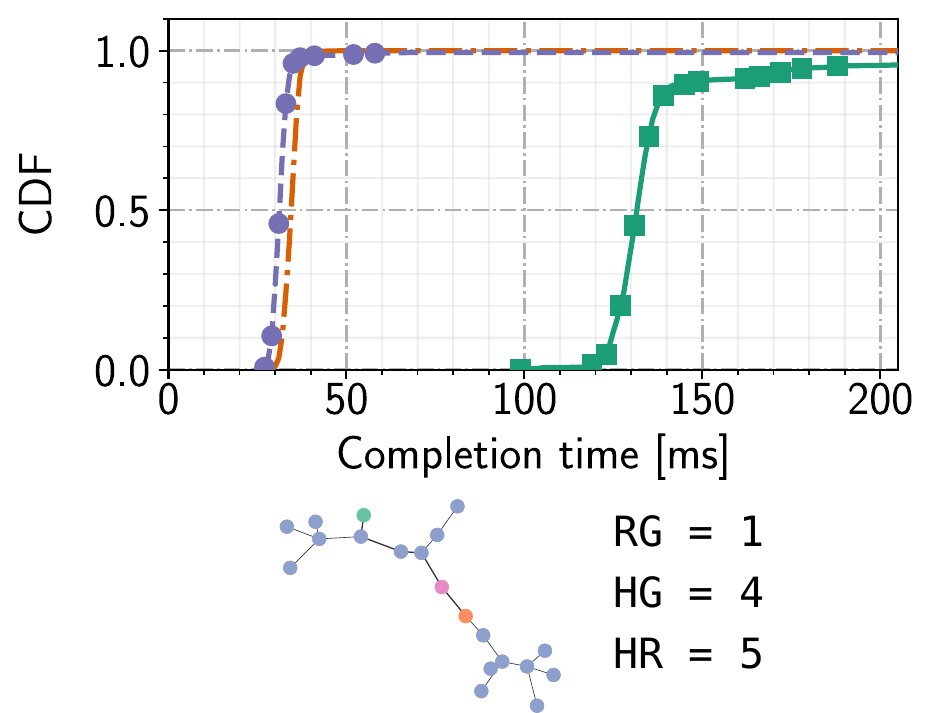}
    \end{subfigure}
    \hfill
    \begin{subfigure}{0.3\textwidth}
        \centering
        \includegraphics[width=\textwidth, height=4cm]{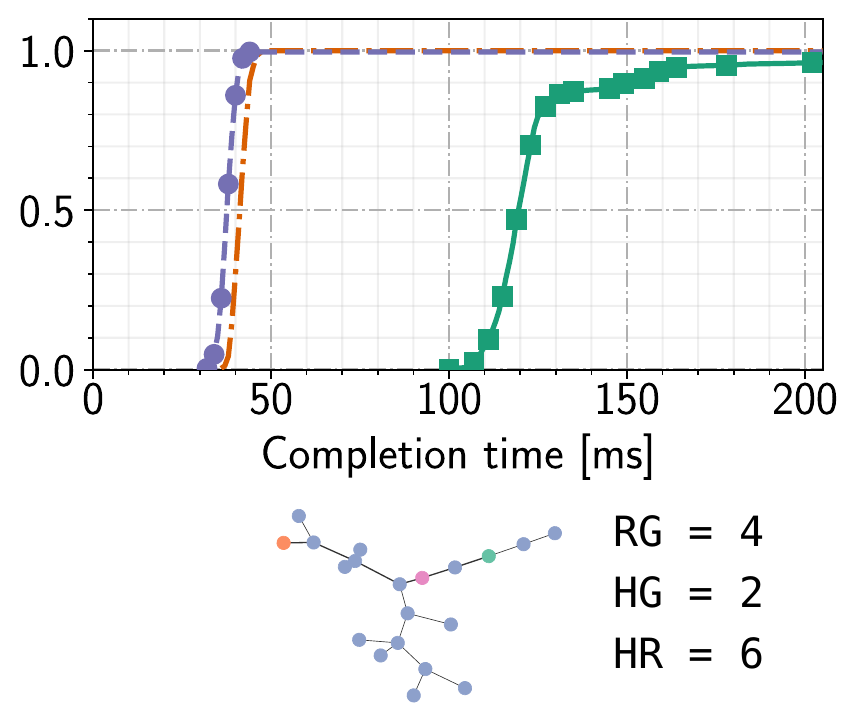}
    \end{subfigure}
    \hfill
    \begin{subfigure}{0.3\textwidth}
        \centering
        \includegraphics[width=\textwidth, height=4cm]{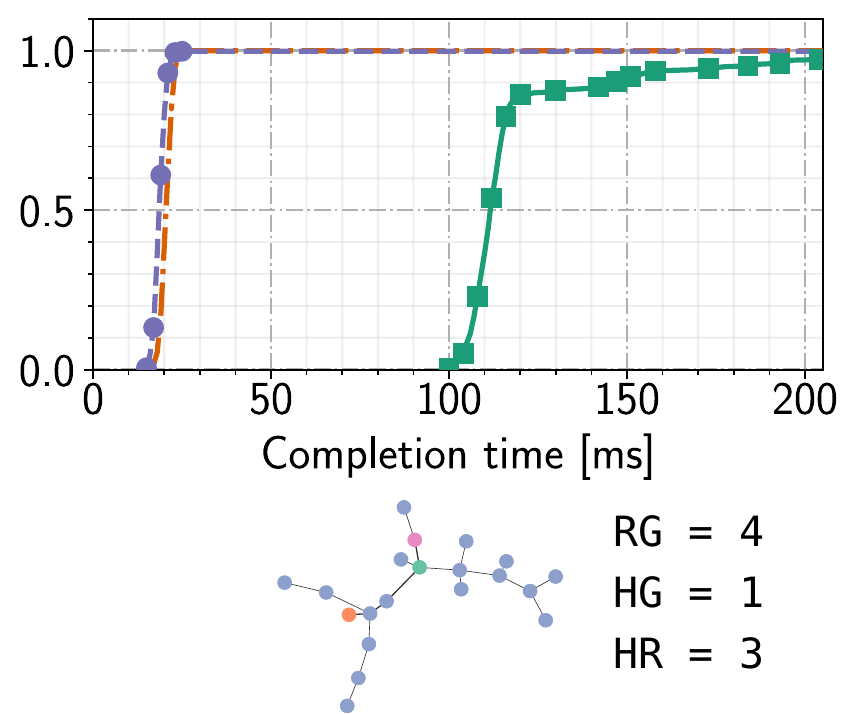}
    \end{subfigure}
    \hfill

    \begin{subfigure}{0.3\textwidth}
        \centering
        \includegraphics[width=\textwidth, height=4cm]{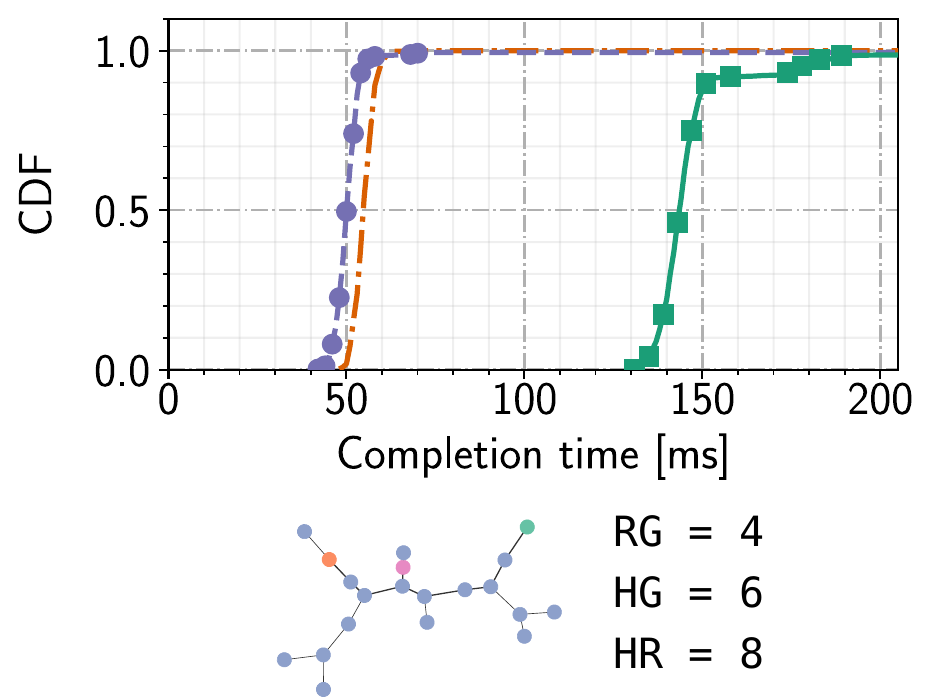}
    \end{subfigure}
    \hfill
    \begin{subfigure}{0.3\textwidth}
        \centering
        \includegraphics[width=\textwidth, height=4cm]{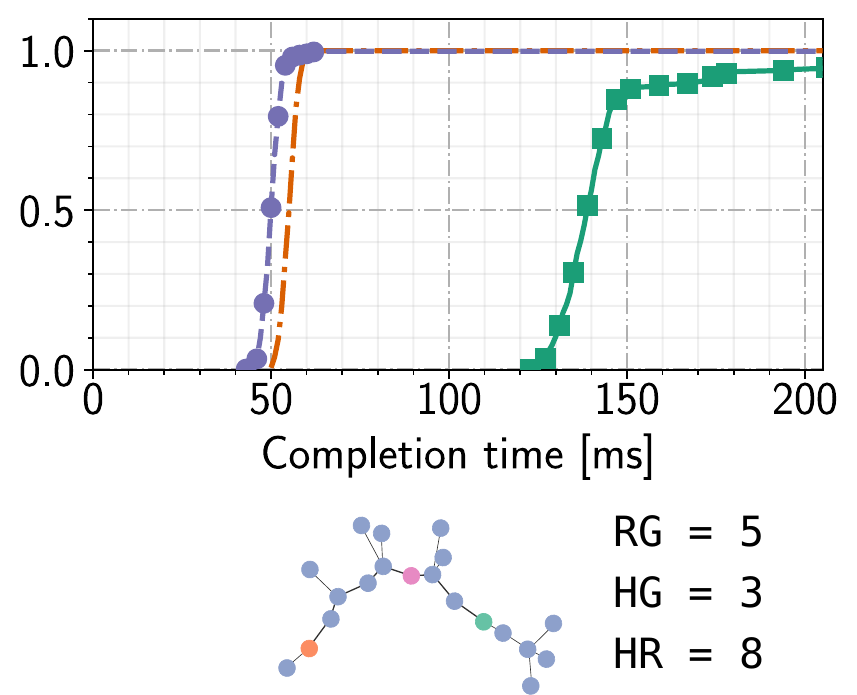}
    \end{subfigure}
    \hfill
    \begin{subfigure}{0.3\textwidth}
        \centering
        \includegraphics[width=\textwidth, height=4cm]{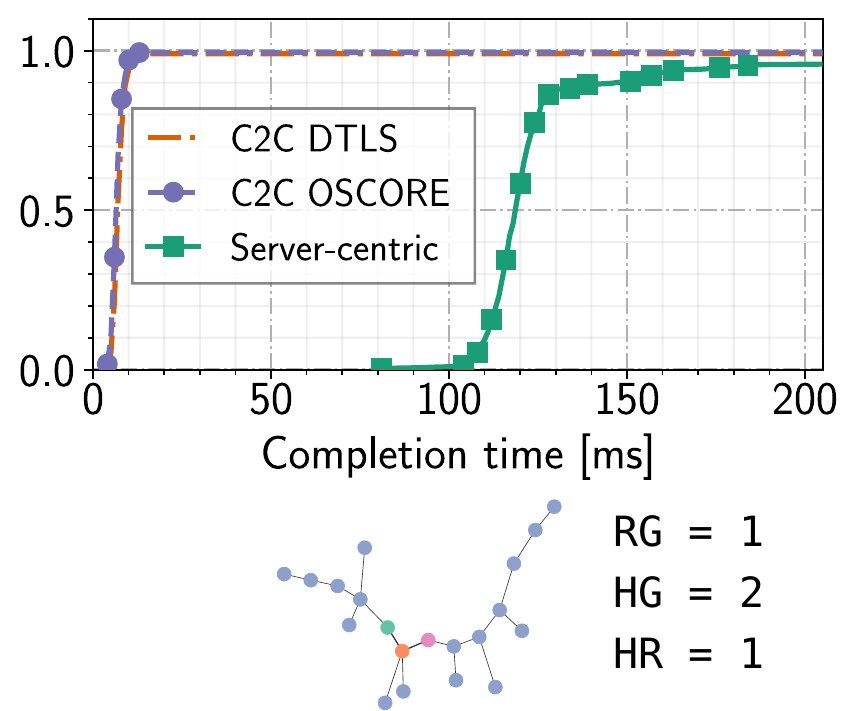}
    \end{subfigure}
    \hfill

    \begin{subfigure}{\textwidth}
        \centering
        \includegraphics[width=0.75\textwidth, height=0.6cm]{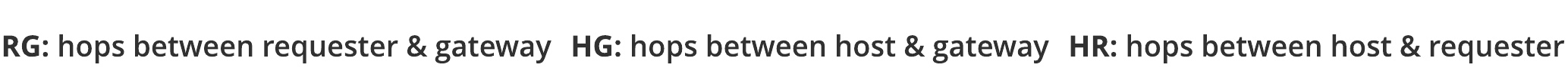}
    \end{subfigure}
    \hfill
    \caption{Temporal distributions of notification arrival times for randomly generated topologies.}
    \label{fig:topologies-notification}
\end{figure*}

\section{Related Work}
\label{sec:related-work}

%
%
\paragraph{Edge computing and device-to-device communication}
IoT deployments have shown a shift from centralized cloud computing paradigms towards distributed architectures that augment the edge of the network \cite{sczlx-ecvc-16} by producing and consuming data locally in an autonomous fashion.
Edge Mesh~\cite{sczy-emnpd-17} proposes a paradigm where decision-making and task distribution are moved to edge devices.
Costa \etal~\cite{cmmp-tltst-06} propose a middleware to share a tuple space among wireless sensors.
Following the same direction, Whitehouse \etal~\cite{wscb-hnasn-04} and Lachenmann \etal~\cite{lmmsg-vsend-07} present neighborhood programming abstractions for wireless sensor networks to share state among them.
Shang \etal~\cite{swabz-ltmr-17} show how named data networking architecture allows building IoT applications with local trust management and inter-vendor interoperability, while staying independent of constant cloud connectivity.
Although these proposals focus on enabling decentralized IoT deployments, they define their own interaction and data models.
In this paper, we focus on LwM2M, as it is a highly deployed management protocol.
Tracey \etal~\cite{ts-haits-13} propose a peer-to-peer architecture based on a distributed hash table, which they further develop in \cite{ts-udppa-19}.
Although they integrate the developed tuple-based library in a LwM2M implementation \cite{ts-lhait-17}, no analysis is performed on the performance impact of such integration on a LwM2M deployment.

%
%
\paragraph{LwM2M extensions}
The interoperability provided by the LwM2M protocol makes it an appealing solution for heterogeneous IoT deployments, thus, multiple extension proposals have been made to expand its capabilities and increase its performance.
Given the lossy nature of IoT networks and the reduced energy availability, there are proposals to reduce the traffic between clients and servers.
Karaagac \etal~\cite{kvrjb-elice-18} define a LwM2M object that allows \server{s} to perform batched operations on clients, thus, reducing the amount of sent messages.
In addition, different LwM2M proxy entities \cite{rdbk-dgmcn-16} \cite{ptm-esllp-20} have been proposed.
They can perform group operations across multiple clients, cache and aggregate responses and apply compression mechanisms to the messages.
We argue that the addition of intermediate proxies in LwM2M networks increases deployment costs and complexity.
Although these solutions reduce the bytes transferred between client and server on certain scenarios, they do not provide direct interaction among clients, thus, they do not enable autonomous deployments with distributed IoT applications,
which is described by Jimenez in \cite{draft-jimenez-t2trg-coap-functionality-lwm2m-00} as a lacking feature in the LwM2M specification.

%
%
\paragraph{Authentication in the IoT}
With an increasing direct communication between constrained IoT devices, the need for mutual authentication and authorization \cite{rfc-7744} arises. Markmann \etal~\cite{msw-feaci-15} propose a lightweight federation scheme that binds device authentication to network attachment. 
Vučinić \etal~\cite{vtrdd-osait-15} propose the OSCAR architecture, based on object security and the distribution of access secrets to request resources from other nodes.
However, the existence of a secret per access group imposes high memory requirements when a fine-grained access control is required. Moreover, the usage of the same secret access across consumers complicates access revocation. AoT \cite{nscnn-aot-16,npsco-abaac-18} proposes a suite of protocols to perform attribute-based access control and authentication throughout the life-cycle of IoT devices. Their analysis suggests that the imposed communication overhead may not be well-suited for limited bandwidth networks, as it would produce a big amount a packet fragmentation, which could lead to packet loss.
Xi \etal~\cite{xqhzz-iraka-16} propose an authentication and key agreement mechanism to enable device-to-device communication, which allows deriving common secrets based on the radio environment. The IETF is developing ACE-OAuth \cite{draft-ietf-ace-oauth-authz}, an authentication and authorization framework based on OAuth 2.0 and CoAP, where devices request access tokens and credentials from authorization servers, which are used to access resources on other devices. This delegates access control and policies to centralized servers.
Although integrating such mechanisms into LwM2M may be feasible, in this paper we focus on reutilizing the already existing key distribution and access control mechanisms in LwM2M.

\section{Conclusion and Outlook}\label{sec:conclusion}

%
%
In this paper, we started from the observation that device management and provisioning is challenging  but required in the constrained IoT.
Popular solutions such as LwM2M avoid this challenge by involving servers when sensors need to communicate with actuators.
This triangular message forwarding requires upstream connectivity where local communication is sufficient.

We designed and implemented client-to-client communication.
Instead of starting completely from scratch, our solution purposefully extends LwM2M.
We provided a detailed security analysis, including an attacker and threat model, which showed that our proposal complies with LwM2M security requirements but abandons the server-centric perspective.
Our performance analysis, conducted in a multi-hop testbed, showed that client-to-client communication leads to shorter data arrival times (up to $\approx$ 90\% on single-hop topologies) and higher and reliable goodput ($\approx 8 \times$ when notifying resource updates) compared to the server-centric communication, but only introduces little overhead ($\approx$ 8\% in ROM and $\approx 0.9\%$ in RAM).
Our findings indicated that client-to-client communication may lead to almost optimal data delivery.

In the future, we plan to apply the principles derived in this work on other management protocols.
We also aim for further improvements of our proposal.
Link Bindings \cite{draft-ietf-core-dynlink} allow to dynamically link state updates between resources, allowing to define distributed behaviours.
By utilizing Group OSCORE \cite{draft-ietf-core-oscore-groupcomm} or its data-centric variants~\cite{gasw-gcorm-21} a \client{} could potentially reduce the number of outgoing notifications when multiple observations exist on the same resources.
The emerging lightweight authorization and authentication framework ACE OAuth may be deployed in parallel to the existing key distribution and access control mechanisms of LwM2M.

\section*{Artifacts}
We conducted all our experiments based on open source software and an open access testbed.
Source code and documentation are available on Github: \url{http://github.com/inetrg/ipsn-2022-lwm2mc2c}.

\begin{anonsuppress}
\begin{acks}
We would like to thank our anonymous shepherd and reviewers for their valuable feedback.
This work was partly supported by the \grantsponsor{BMBF}{German Federal Ministry of Education and Research (BMBF)}{https://www.bmbf.de/} within the project \grantnum{BMBF}{PIVOT}.
\end{acks}
\end{anonsuppress}

\balance
\bibliographystyle{ACM-Reference-Format}
\bibliography{meta,rfcs,security,own,theory,ids,ngi,iot,programming}

\label{lastpage}

\end{document}